\documentclass[namedreferences,psfig]{kluwer}

\usepackage{epsfig}

\def\msol{{M}_{\odot}}
\def\vsol{{v}_{\odot}}
\def\eros{{\sc eros}}
\def\erod{{\sc eros2}}
\def\erou{{\sc eros1}}
\def\macho{{\sc macho}}
\def\ogle{{\sc ogle}}
\def\gman{{\sc gman}}
\def\mps{{\sc mps}}

\def\planet{{\sc planet}}
\def\lmc{{\sc lmc}}
\def\smc{{\sc smc}}
\def\ie{{\em i.e.}}

\begin{document}
\begin{article}
\begin{opening}

\title{The Galactic Halo from Microlensing}
\runningtitle{The Galactic Halo from Microlensing}

\author{Alain \surname{Milsztajn}\email{AMilsztajn@cea.fr}}
\runningauthor{A. Milsztajn}

\institute{DSM--DAPNIA--Service de Physique des Particules, 
CEA-Saclay, F-91191--Gif-sur-Yvette, France}

\begin{abstract}

The status of the microlensing search for galactic dark matter in the
form of massive astronomical compact halo objects
(machos) is reviewed.  Unresolved issues are 
discussed, as well as possible ways to solve these. 

\end{abstract}
\end{opening}

\section{Introduction}

The mass of our Galaxy can be computed from dynamical studies of 
its rotation and of the motion of its satellites, or alternatively
it can be evaluated from its visible components, primarily stars.
That these two estimates disagree by a factor 5-10 constitutes
the problem of Galactic dark matter.  Either the laws of dynamics 
we use are wrong on galactic scales (few kiloparsecs), or there exists 
some form of galactic matter that does not emit or absorb
enough electromagnetic radiation to be directly detectable.  
Studies of many other spiral galaxies confirm that this problem
is not unique to the Milky Way.  

Originally proposed by B.~Paczy\'nski (1986) as a probe of galactic
dark matter, the gravitational microlensing technique relies on
detecting the transient magnification and/or deflection 
of the light from extragalactic stars by intervening machos. 
The a priori mass range for machos is very wide, between about
$10^{-7}$ and $10^4 \:\msol$ (corresponding to event durations ranging
from one hour to a few decades). 
Lighter primordial H/He objects would 
have evaporated since the galaxy formed \cite{ruj92}; a halo full of 
heavier objects would be devoid of globular clusters by now
\cite{arr99}. 

The search for gravitational microlensing phenomena in our Galaxy
has now been going on for over a decade.  
Although the prime suspects for machos were initially 
brown dwarfs (0.01 to $0.1~\msol$), 
and although many arguments existed against 
real or imaginary objects of other masses, the survey groups have
chosen to cover the widest possible fraction of the 11 orders of
magnitude macho mass range.   The first microlensing 
candidates were reported in 1993, towards the \lmc\ 
\cite{alc93,aub93} and the Galactic Centre \cite{uda93}
by the \eros , \macho\ and \ogle\ collaborations.  As the present
review centers on halo dark matter, I will mostly discuss results
from the former two groups. (\ogle\ microlensing studies have
concentrated on the Galactic Bulge, see e.g. \opencite{uda00}.)

Other groups contribute to the field. Three of them 
are follow-up groups, {\sc gman, mps} and {\sc planet}; they observe
microlensing events alerted upon by the survey groups.
The {\sc moa} group conducts a survey mainly dedicated to the
search for planets (see e.g. \opencite{moa01}), a topic that is also
a prime motivation of {\sc mps} \cite{mps00}
and {\sc planet} \cite{pla00}.  Finally, the {\sc point-agape} group
attacks the more ambitious goal of microlensing
towards M31 \cite{aga01}.
 
As the present review is short and contains no calculations, 
I refer the interested reader to other sources 
\cite{pac96,rou97,gou00,gou01}.

This article is organised as follows.
In Sect.~2, I recall the main properties of microlensing and the
information that can be obtained from its observation. 
Sect.~3 discusses results from observations towards the Large
Magellanic Cloud (\lmc ).
In Sect.~4, the same is done for the Small Magellanic Cloud (\smc ). 
Sect.~5 contains a critical assesment of these results, and 
conjectures about what the future may hold.

Finally, Sect.~6 gives my conclusions.

\section{Properties of Microlensing}

\subsection{Simple Microlensing}

When both the source star and the lens are simple (non-binary) 
and their size can be neglected, microlensing
depends on two distances and one mass \cite{ein36}.  
The magnification is
$ A(u) = (u^2 + 2) / [ u^2 (u^2 + 4)]^{1/2} $, where $u$ is the ratio
$ \theta / \theta_E $. Here, $\theta$ is 
the angular separation between the lens and the ``true'' position
of the source (\ie\ when the lens is far away), and  $\theta_E$ is
the natural angular scale for significant microlensing,
$ \theta_E = [ 4 G M / c^2 \cdot (1-x) / Lx ]^{1/2} , $ where $M$ is the lens
mass, $Lx$ its distance and $L$ the distance to the source\footnote{ 
$\theta_E$ is also the angular radius of the so-called ``Einstein'' 
ring occuring for $\theta = 0$.}.  For small $u$, $A \simeq 1/u$;
at large $u$, $A$ dies out quite fast, $A \simeq 1 + 2 / u^4$.

As the lensing object moves with respect to the line of sight to the
source star, microlensing phenomena are transient. Their natural 
timescale is the time needed to move by an angle equal to $\theta_E$,
$t_E = \theta_E / \mu $, where $\mu$ is the proper motion (angular
velocity) of the lens w.r.t. the line of sight. The simplest 
microlensing light curves depend on four parameters, the baseline 
(unmagnified) flux, the time $t_0$ of maximum magnification $A_{max}$, 
and the timescale $t_E$. If a significant fraction of the flux comes
from a blended star many $\theta_E$ away, a (fifth) blending parameter
is necessary. 

These four (five) parameters can be determined by fitting the known
functional form of the light curve to the measurements.  Of these
parameters, only one contains information on the lens, $t_E$, and 
this is degenerate in the lens mass, distance and velocity. Two 
other parameters can be used, in a statistical sense, to test that
the observed light curves are indeed due to microlensing~: the source
star magnitude and the minimum impact parameter $u_0$
($A_{max} = A(u_0)$). If one can ignore detection efficiency and
blending effects, $u_0$ should display a flat distribution; its true
expected distribution can be obtained from a simulation of the 
observing program.  Similarly, as the lens does not choose the source
it lenses, the source star should be distributed randomly in the
colour-magnitude diagram, once again apart from detection efficiency
which favors brighter stars. At present, such tests have only been 
done convincingly for microlensing in the galactic plane, where
enough events have been found (few hundreds).

Ignoring efficiency, the average expected microlensing timescale 
for Magellanic Cloud stars is 
       $$t_E = 70 \, \mathrm{days} \cdot (M / \msol)^{1/2} . $$ 
(This is obtained 
from an ``isothermal'' spherical halo model that explains the galactic
rotation curve, with $4 \, 10^{11} \msol $ within 50~kpc.) The expected
average $\theta_E$ is about $0.8 \, \mathrm{mas} \cdot (M / \msol)^{1/2} $,
which explains why the search for microlensing has concentrated up to 
now on observing the magnification, not the deflection.  Astrometric
microlensing (\ie\ measuring the photocentre motion during microlensing)
will be feasible quite soon, especially with the foreseen {\sc fame}, 
{\sc gaia} and {\sc sim} astrometric satellites. 
I do not discuss it further and refer the reader to the review of
\inlinecite{gou01}; let me mention only that astrometric microlensing
has a shallower dependence on the impact parameter, and that it 
provides information complementary to that of photometric microlensing;
this will help in breaking the timescale degeneracy. 

Apart from microlensing timescale, the other important observable is
its rate.  As $\theta_E \propto M^{1/2}$, a dark lens produces significant
microlensing in a solid angle $\propto M$, and the probability that 
microlensing is currently occuring
on a given star is thus proportional to the (weighted) integral of the
mass density along the line of sight.
This is known as the ``optical depth'' and it can be shown to be
of order $(\vsol / c)^2$ ($\vsol$ is the galactic rotation
velocity). This very low probability is the key factor for 
the observing strategy of the survey groups~: they must monitor
few tens of million stars a few times per week in order to reach 
the necessary sensitivity.
Experimentally, the optical depth 
is measured from the detected microlensing events as 
$$ \tau = \frac{\pi}{2} \frac{1}{N_* T_{obs}} 
\sum_{evts}\frac{t_E}{\epsilon(t_E)} , $$
where $N_*$ is the number of stars monitored for a duration $T_{obs}$,
and $\epsilon(t_E)$ is the overall detection efficiency for events with 
timescale $t_E$.\footnote{The definition of $\epsilon$ here is such that
$\epsilon$~=~100\% if all microlensing events with $u_0 < 1$ occuring
within $T_{obs}$ are detected.}

Finally, when blending is negligible or is due to a star with the same
colour as that being microlensed, microlensing is achromatic.  Survey
groups thus monitor their fields in at least two bandpasses, using this 
as a tool to identify intrinsic variable stars, that are at least 
$10 \, 000$ times more abundant than microlensing phenomena.   

\subsection{General Microlensing}

In order to learn more about the lensing object, one needs more
observables than just $t_E$.  This is possible when the approximation
of simple microlensing breaks down, and when another distance scale
(perpendicular to the line of sight) 
can be measured and compared to $\theta_E$.  There are three classes
of such scales, according to whether they lie in the observer, source
or lens plane.

In the observer plane, the Earth orbital motion can lead to measurable
distortions that allow one to compare its semi-major axis to $\theta_E$
for long enough events, $t_E > 3$~months \cite{gou92}. 
Such a ``parallax'' distortion is largest when the lens is closer
to the observer, and unobservable when it is located near the source,
in the Magellanic Clouds.  Alternatively, the same microlensing 
phenomenon observed simultaneously from two 
distant points in the solar system could provide the same information
\cite{ref66}
(see \inlinecite{ogl43} for a recent spectacular illustration of these).
As they provide an additional distance scale in the observer plane,
such situations allow one to measure the angular velocity of the lensing
object as would be seen from the source star, $\mu_{MC} = \mu x / (1-x)$.  
This provides some discriminating power~: Magellanic Cloud lenses 
will have a high MC-based angular velocity $\mu_{MC}$, 
whereas halo lenses will generally show a lower value.

In the source plane, the separation of the two components of a binary
source star can play the same role, provided their revolution period
is measured; examples are events
{\sc eros2-gsa2} \cite{der99} and {\sc macho-96-lmc-2} \cite{alc01}.
Alternatively, the source star diameter can be used, especially when 
it nears or crosses a caustic curve (see e.g. \opencite{afo00}
and references therein).  
In the case of a simple lens, the caustic ``curve'' reduces to 
a point, $\theta = 0$.  For a double lens, caustic curves have a
more complicated structure, that can however be determined from 
the mesured light curve.
If the source star angular diameter is known, the time it takes to
cross the caustic curve is a rather direct measure of the relative 
proper motion of the lens, $\mu$, relative to the line of sight.

In the lens plane, finally, a double lens with measured separation 
and period would also allow to reduce, or even break the degeneracy.

Another interesting observable regarding double (or multiple) lenses
is the relative rate of these with respect to simple microlensing.
Double lenses can be detected as such when the component angular 
separation is of order $\theta_E$.  When they are much closer, they act
effectively as a single lens; when they are much further, one component
is simply not detected from the light curve. The rate for detectable 
double lensing was estimated at 5-10\% for galactic plane lenses
\cite{mao91, gri92}. Recent measurements by \ogle\ \cite{uda00}
bear out this prediction.

\section{Observations towards the LMC}

The Magellanic Clouds are close enough that many tens of million 
stars can be resolved from comparatively small ground-based telescopes,
and far enough that a search for microlensing will sample an 
appreciable fraction of the galactic halo.  Hence, they are  
presently the main targets in the search for machos.  
I first discuss observations towards the \lmc .

Because they observed no microlensing candidate with a timescale
shorter than 10~days, down to 0.04~day (1 hour),
the \erou\ and \macho\ groups were able to exclude 
the possibility that more than 10\% 
of the Galactic dark matter resides in planet-sized objects
($10^{-7}$ to a few $10^{-3} \msol$)\footnote{Slightly  
weaker limits were
obtained for heavier machos of about $10^{-2} \msol$.}.  
  This resulted from a dedicated program by \erou\ 
\cite{aub95,ren97,ren98}, 
or as a by-product of the search for heavier machos  
\cite{alc96}.  The two groups have combined their
limits in \inlinecite{alc98}.  

In contrast, 
the first few events that were detected with longer time\-scales
aroused at first high hopes that the Galactic dark matter
problem was about to be solved.   
In the analysis of their first two-year data set \cite{alc97a}, 
the \macho\ group found
6-8 candidate events towards the \lmc\ out of 
7.5 million monitored stars.  They estimated  
an optical depth of
order half that required to account for the dynamical mass of a 
``standard'' spherical dark halo\footnote{
$4 \times 10^{11}\:{\rm M}_\odot$ within 50~kpc};
the typical timescale of the events, $t_E$, 
implied an average mass of about $0.5 \msol$ for halo lenses. 
The statistics was small enough that a halo
full of machos was compatible with the result.
Based on two candidates out of 3.3 million monitored stars, 
\erou\ preferred to set an upper limit on the 
halo mass fraction in objects of similar masses
\cite{ans96,ren97}.  For typical brown dwarf masses, the limit was 
below that required to explain the rotation curve of our Galaxy, 
but it did not conflict with the \macho\ result.

The first results towards the \smc\ then appeared (see next section).  
The \erod\ group published an even lower limit on machos, now
excluding the upper part of the domain allowed by the \macho\
results.  It then seemed less likely that the halo could be fully 
comprised of $0.5 \msol$ machos.

The situation was partly clarified in 2000. 
The \macho\ group presented an analysis of 5.7 year light curves 
of 10.7~million stars in the \lmc\ \cite{alc00} with an
improved determination of their detection efficiency and a better
rejection of background supernova explosions behind the \lmc .
A total of 13-17 microlensing candidates were found (including
the previous 6).  These two numbers refer to two different 
analyses of the same data, with stricter or looser selection
criteria. These results from the \macho\ group
now favour a galactic halo macho component of 20\% in the
form of $0.4 \msol$ objects.  This is their central value,
but their result is compatible, at 95\% C.L., with a halo fraction
ranging from 8 to 50\% and a macho mass between 0.15 and $0.9 \msol$ 
(see Fig.~1).  At about the same time,
the detection of a halo white dwarf population
at the level of a 10\% component was also reported by 
\inlinecite{iba00}. 

Almost simultaneously, the \erod\ group
presented its results from a two-year survey of 17.5 million stars 
in the \lmc\ \cite{las00a}.  
One of the two \erou\ microlensing candidates 
was seen to vary again, 8 years after its first brightening, and
was thus eliminated from the list of microlensing candidates.
Two new candidates were identified. 
A few months later, this analysis was extended to three-year
light curves of 25.5 million stars; this essentially confirmed
the results of the two-year analysis \cite{mil01}.  
The total number
of microlensing candidates towards the \lmc\ from \eros\ is now
five, one from \erou\ and four from \erod .  
The candidates timescales agree with those from \macho , at an 
average of about 30~days.  Provided the events are due to halo 
lenses, this is compatible with $0.4 \msol$ lensing objects.
The rate measured by \eros\ is about twice lower than that of \macho . 
Because this rate is 10 times lower than expected if \macho s
are a substantial component of the galactic halo, and because the
four new \eros\ candidates do not show excellent agreement with simple
microlensing light curves, \erod\ continues to 
quote an upper limit on the fraction of the galactic halo 
in the form of \macho s (see Fig.~1). It can be seen that this
combined limit from all \eros\ data sets excludes the upper 25\% 
of the parameter space domain favored by \macho .

%
%
\begin{figure} [ht] 
\epsfig{file=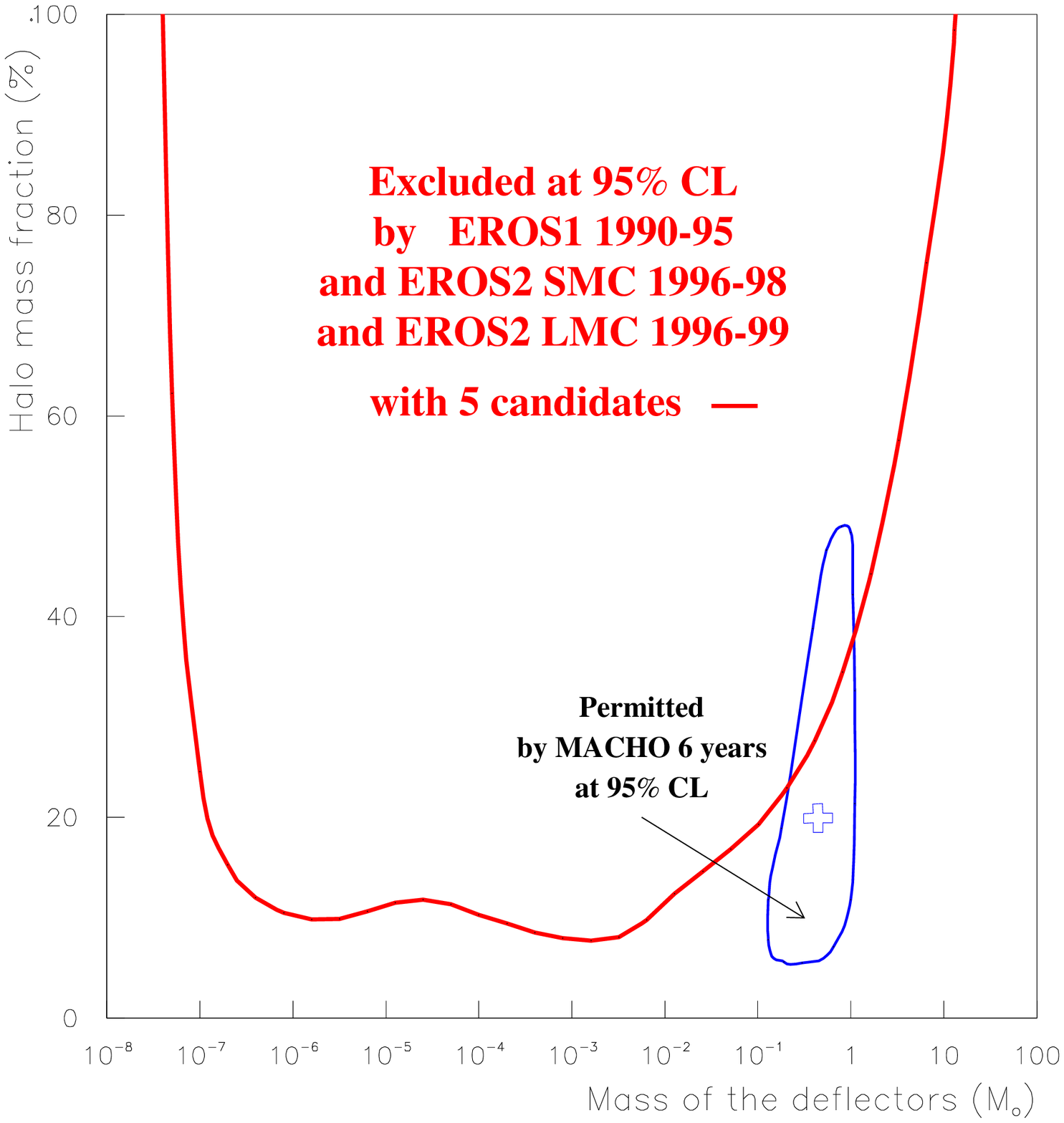,width=12cm,clip=}
  \caption[]{Results from {\sc eros} and {\sc macho} 
  on the macho component of the galactic halo, between 0 and
  100\% of a standard halo model ($4 \times 10^{11}\: \msol$ 
  inside a 50~kpc radius sphere), for machos with masses ranging
  between $10^{-8}$ and 100~$\msol$.
  The solid line is the 95\% C.L. upper limit inferred by \eros\  
  from their five \lmc\ microlensing candidates found in all
  \lmc\ and \smc\ data, 1990-99 \cite{mil01}.
  The closed contour is the \macho\ 95\% C.L. accepted region, 
  from their 13 \lmc\ microlensing candidates found in 1992-98
  \cite{alc00}.
  The preferred value is indicated by the cross.} 
\end{figure}
%
%

The 13 \macho\ microlensing candidates selected with stricter criteria
all have timescales smaller than 75~days.  The \macho\ group has used
this to obtain an upper limit on black hole dark matter in the
galactic halo \cite{alc01a}.  They conclude that less than 40\% of the
halo dark matter can be in the form of 1 to $10 \, \msol$ objects. 
Altogether, microlensing surveys have now probed 
slightly more than 8 orders of magnitude in possible macho masses.
  
Some microlensing events towards the \lmc\ display interesting
peculiarities.  Event {\sc macho-lmc-1} is much better fit by a
binary lens with no caustic crossing \cite{dom96,alc00a}; this is 
presently the shortest timescale candidate towards the Magellanic 
Clouds. Event {\sc macho-lmc-9} is a double lens with caustic 
crossing \cite{alc00a}; its proper motion is very low, thus favoring
an interpretation as a double lens within the \lmc \footnote{This
view is not completely shared by \inlinecite{alc00a}.}.  The source star for
event {\sc macho-lmc-14} is double \cite{alc01} and this has 
allowed to conclude that the lens is in the \lmc . 
Event {\sc macho-lmc-5} is probably due to a galactic disk lens 
(\inlinecite{gou97} and 
D. Bennett, 2001, personal communication).

\section{Observations towards the SMC}

Although the \smc\ offers about 5-10 times less stars than the \lmc\
for microlensing studies, it is a valuable target.  
Because the lines of sight to the \lmc\ and \smc\ are close 
(about 20~degrees apart), 
the timescale distributions of microlensing candidates towards
the two Clouds should be nearly identical if lenses belong to
the galactic halo. (The difference in the average timescale for
lenses of a given mass is much lower than the width of the 
distribution.)  This point has been emphasised by \inlinecite{gra01}
-~as reported in \inlinecite{mil01}~- as a way to distinguish between
microlensing being due primarily to
halo lenses or to Magellanic Cloud lenses.

Also, the event rates should be comparable, 
although the ratio is more halo model dependent. 
Of course, this model dependency can be turned into an advantage,
and may allow one to estimate the flattening of the halo 
\cite{sac93,fri94,alc95},
if this is where the lenses are. 

At the time of this review (June 2001), only
two microlensing events have been
detected towards the \smc , {\sc macho-97-smc-1} 
($\equiv$ {\sc eros-97-smc-1}) and
{\sc macho-98-smc-1}.  These events imply an 
optical depth towards the \smc\ which is consistent with that measured
towards the \lmc\ by the \macho\ group. 
(Because of the small statistics, this is not very informative.) 
However, further information exists on both events that 
point to microlensing due to \smc\ lenses.

The situation regarding the second microlensing event,
{\sc macho-98-smc-1}, is straightforward.  
It was first alerted upon by the
\macho\ group on 25 May 1998, and a probable caustic crossing due to 
a double lens was announced soon after, on 8 June 1998.
The source star was monitored by most microlensing groups.
This allowed the second caustic 
crossing to be entirely covered, by the \planet\ and \eros\ 
groups, resulting in a measurement of the crossing duration, from
which the relative proper motion could be measured.  
Separate or combined analysis of the data from all five
groups led to the conclusion that the double lens
proper motion, about 1.4 ~km~s$^{-1}$~kpc$^{-1}$,
is extremely unlikely to be that of a halo object
(\opencite{afo00}, and references therein).  
Therefore it must be located in the \smc .

The first candidate microlensing event towards the \smc\
was reported by \inlinecite{alc97b}.  
It was observed independently by \inlinecite{pal98}.
As observed by \inlinecite{pal98} and confirmed by \inlinecite{uda97},
the source star is a periodic variable (P~=~5.12~d; A~=~5\% ). 
It is bright, actually the brightest for all microlensing
events towards the Magellanic Clouds.
It also has the longest timescale of all \lmc\ or \smc\ events,
$t_E \simeq 125$~d.  As such, it is a particularly good place 
to look for a ``parallax'' distortion (see Sect.~2.2).

No parallax could be measured from \erod\ data,
indicating that the lensing object is probably in the \smc\ 
\cite{pal98}. 
\inlinecite{gra01} have also searched for a parallax distorsion
in the \macho\ data for this event; this unsuccessful search 
produces a lower limit on $\mu_{MC}$ that strengthens this
conclusion, at the 97\% C.L. 

The present situation is thus that none of the two microlensing
events observed towards the \smc\ are due to halo lenses.
The \erod\ group has used this to present limits obtained from
their \smc\ two-year data alone on the fraction of stellar mass
machos in the halo \cite{afo99}.

\section{Discussion and Prospects}

I now discuss a few topics arising from these results in more detail.

\subsection{Statistical Tests} 

The {\sc macho} collaboration favours an interpretation 
of their candidates in terms of galactic halo lensing objects.  
They find that their sample is not, or little, contaminated
by variable stars.  In support of this, 
the distribution of stellar luminosities of their 
microlensing candidates agrees with that of \lmc\ stars 
\cite{alc00, alc01c}. 
The distribution of maximum magnifications 
is also compatible with a random minimum distance of approach of the
lens to the star line of sight.  These two tests could have revealed
a possible contamination of the sample by intrinsic variable stars.
(The same is true for the \eros\ candidates, but the small number of 
candidates precludes any significant conclusion.) 

\subsection{MACHO vs. EROS : Optical Depth, Blending and Efficiency } 

Even if all \eros\ candidates are {\it bona fide} microlensing 
phenomena, the corresponding optical depth is twice lower than
that from \macho .  As the two numbers are compatible within errors,
this might simply be a statistical fluctuation.  It seems useful,
though, to search for candidates that should have been observed 
by both surveys. Such a study reveals no problem.  
The actual overlap between the two data sets is not
very large, and only one candidate is found in both analyses~:
candidate {\sc eros2-lmc-5} is identical to {\sc macho-lmc-26}, which 
does not meet the \macho\ strict selection criteria.
Recall also that the first \smc\ microlensing event was observed 
by the two groups. 

An accurate measurement of the optical depth requires knowledge
of the number of monitored stars $N_*$, and unbiased estimates
of the ratio $t_E / \epsilon(t_E)$ for all candidate events.
Here, the effect of blending can be serious~: it leads to an
underestimate of $t_E$, and of $A_{max}$ as well. (The latter
can influence the candidate list through a minimum cut on this
variable.)  An underestimate of $t_E$ however can be partially 
compensated by the corresponding underestimate of the detection 
efficiency, $\epsilon$.  On the other hand, faint stars blended
with brighter ones can lead to additional detectable microlensing,
for low enough impact parameters $u_0$.  This can best be 
corrected for through simulated images.  On that topic, 
the \macho\ treatment is 
in principle more complete than that of \eros , and this could be
an important point if the blending correction is large.  But
\eros\ fields are slightly less crowded in average, and studies have 
indicated that the overall effect of blending is small.
Note that efficiencies from \macho\ are larger than those of 
\eros ; this depends a lot on the list of monitored stars, the 
sampling and on the chosen analysis cuts.  A larger
correction for blending in the \eros\ data would lead to a 
{\it larger} efficiency, and thus a lower microlensing optical depth,
thus increasing the difference with \macho .
 
Could it be that the different fields followed by \eros\ and 
\macho\ really have different optical depths~?  This would certainly
be the case if most lenses are in the Magellanic Clouds, as the 
optical depth would then be larger towards the centre, where there 
are more stars, \ie ~more lenses.  The importance of this so-called
``self-lensing'' was first stressed in \cite{wu94,sa94}.
\eros\ covers a larger solid angle in the \lmc\ (64~deg$^2$)
than \macho , which monitors primarily
15~deg$^2$ in the central part of the \lmc .
The \eros\ rate should thus be less contaminated by ``self-lensing''.
However, the {\sc macho} group has compared their sample to 
predictions of a specific model of the \lmc , and
finds that the observed distribution favours halo lenses, although 
they cannot completely exclude \lmc\ lenses in a kind of 
\lmc\ ``halo'' \cite{alc00}.  

\subsection{Localised Lenses} 

Out of the 20 or so microlensing candidates towards the 
Magellanic Clouds, it has been possible to localise with reasonable
certainty five of them.  None of these localised lenses are
in the halo.  The two \smc\ events are due to \smc\
lenses (see Sect.~4), one from its binary lens nature, the
other one from the absence of a measurable microlensing parallax.
Three \lmc\ event lenses have been localised, two in the \lmc\ 
(one double lens, one double source) and one in the galactic disk.

The Magellanic Cloud double lenses ({\sc macho-98-smc-1} and
{\sc macho-lmc-9}) are especially interesting.  If one 
naively extrapolates to Cloud lenses the result that, 
for galactic disk lenses, about 10\% show their binarity in 
the light curve \cite{uda00}, 
one is tempted to ask where are the 20 
($= 10 \times 2$) simple Magellanic Cloud lenses.  
Here again, the small statistics forbids us to go much further.

\subsection{More Statistics ? } 

What are the prospects for more statistics towards the \lmc\ ?
The \macho\ group has analysed 5.7 years of data out of a final
total of 7.5 years (observations were terminated at the end of 1999).
This would represent only a modest increase.  
In addition to the 15~deg$^2$ analysed in \inlinecite{alc00}, 
25 more have been monitored with a less frequent sampling.
These fields would have a lower efficiency for the range of $t_E$ 
displayed by present candidates, but would increase the 
sensitivity to heavier machos.  The \eros\ group has analysed 
3 years of data for 25 million stars, and expects to have a final
data sample (by 2002) of 35 million stars over 6 years.  This would
represent an almost threefold increase in sensitivity.  Also, the 
\ogle\ group has recently commissionned a new camera that will 
allow them to multiply their data taking rate by a factor 5;
they will probably follow more Magellanic Clouds fields 
than in the past.

\subsection{Comparing the LMC and SMC candidates} 

What are we to think of the present number of halo lenses (zero) 
in the search for microlenses towards the \smc\ ?  In other
words, how many \smc\ halo microlens events (with 
$t_E^{SMC} \simeq t_E^{LMC}$) should have been observed by both groups, 
if the signal presented by \macho\ is entirely due to halo lenses~? 
\inlinecite{gra01} have investigated this. 
The \macho\ group monitored at least 2.2 million stars
and has observed them over 6.5 years (Alcock et al. 1997b).
Assuming equal detection efficiencies for \smc\ and \lmc\ 
target stars, \citeauthor{gra01} estimate that \macho\ should 
observe in average $3.0 R$ events, where 
$R$ is the ratio of the optical depth towards the \smc\ and \lmc .
In the standard spherical halo model, $R = 1.4 \:$, and thus the
expected number of events is about 4.2.
In the same way, using the published \eros\ efficiencies towards
the \smc\ \cite{afo99}, one obtains for 6 years monitoring of
5.3 million stars about 3.6 events. \citeauthor{gra01} estimate
that, out of these $4.2 + 3.6 = 7.8$~events, only about 1.1 are
common.  Hence, provided none of the above assumptions is wrong,
the prospect of an independent measurement of the optical depth 
towards the \smc\ looks good.  Agreement in the microlensing 
rates and durations would represent, in my opinion, one of the
most interesting tests of the halo lens hypothesis.     

\subsection{Halo White Dwarfs} 

What could the $0.4 \:\msol$ objects implied by the \macho\ 
result be~?  A possible answer is white dwarfs, although the mass
seems a bit low.  Many groups have undertaken the search for 
halo white dwarfs, that would be detected directly through 
their high proper motion when they traverse the galactic disk 
in the solar vicinity.  A detection has been presented by 
\inlinecite{iba00}, at the level of 10\% of the expected 
local dark halo density.  Other surveys report negative results 
\cite{fly00, gol00}.  More recently, another detection was reported 
at the level of 2\% of the expected local dark halo density 
\cite{opp01}.  This result has been  
criticized by \inlinecite{gra01b}, \inlinecite{gib01},
\inlinecite{rei01} and \inlinecite{rey01}.
At present, it does not seem possible to identify 
these possible detections with the objects responsible for 
the \macho\ result.

\section{Summary }

Microlensing surveys have not (yet ?) found enough machos to
explain the galactic rotation curve.  The undisputed result 
of the \eros\ and \macho\ groups is that
there exist strong limits on machos in the mass range between 
$10^{-7}$ and $10^{-1} \msol$.  Between 0.1 and 
$1 \:\msol$, the two groups have candidates, some of which are
undisputable microlensing phenomena.  Their results are compatible,   
as is apparent in Fig.~1, but they interpret them differently.
In my view, the issue of where the lenses are is not yet solved, 
but there are good prospects for doing so in the coming years.
No long timescale candidates have been seen, that could be 
attributed to a halo object more massive than the Sun.
Between 1 and $30 \:\msol$, the sensitivity of the two surveys 
is such that they should have already seen a few events
due to putative heavy machos; the ensuing limits are still weak.
Above $30 \:\msol$, there is not enough sensitivity yet.  In the 
absence of alternative ideas to look for such heavy machos, 
any progress from the surveys in this mass range is welcome.

Finally, the search for microlensing towards the Magellanic 
Clouds will not stop with \eros\ and \macho .
Very soon, larger earth-based surveys, such as {\sc ogle3}  
or {\sc supermacho}, will also contribute to this topic. 
Microlensing towards M31 will provide a new line of sight that
should help in interpreting the results.  A few years later,
simultaneous astrometric and photometric microlensing 
measurements from astrometric satellites will allow 
to directly find the lens position, for those few events with
measurable ``microlensing parallax''.  

\begin{acknowledgements}
I would like to thank D.~Graff, T.~Lasserre and J.~Rich for their 
comments and careful reading of the manuscript.
\end{acknowledgements}

{}

\end{article}
\end{document}